# Evidence of thermal transport anisotropy in stable glasses of vapour deposited organic molecules

Joan Ràfols-Ribé[1], Riccardo Dettori[2], Pablo Ferrando-Villalba[1], Marta Gonzalez-Silveira[1], Ll. Abad[3], Aitor Lopeandía[1], Luciano Colombo[2], Javier Rodríguez-Viejo[1,*]

[1]Nanomaterials and Microsystems Group, Physics Department, Universitat Autònoma de Barcelona, 08193 Bellaterra, Spain.

[2]Department of Physics, University of Cagliari, Cittadella Universitaria, 09042 Monserrato (Ca), ITALY.

[3]IMB-CNM-CSIC, Campus Bellaterra, 08193 Bellaterra, Spain

Abstract

Vapour-deposited organic glasses are currently in use in many optoelectronic devices. Their operation temperature is limited by the glass transition temperature of the organic layers and thermal management strategies become increasingly important to improve the lifetime of the device. Here we report the unusual finding that molecular orientation heavily influences heat flow propagation in glassy films of small molecule organic semiconductors. The thermal conductivity of vapour-deposited thin-film semiconductor glasses is anisotropic and controlled by the deposition temperature. We compare our data with extensive molecular dynamics simulations to disentangle the role of density and molecular orientation on heat propagation. Simulations do support the view that thermal transport along the backbone of the organic molecule is strongly preferred with respect to the perpendicular direction. This is due to the anisotropy of the molecular interaction strength that limit the transport of atomic vibrations. This approach could be used in future developments to implement small molecule glassy films in thermoelectric or other organic electronic devices.

*Corresponding author: javier.rodriguez@uab.es

**I. INTRODUCTION**

Organic glasses are important for a wide range of scientific and technological processes [1]. Their utilization in organic electronics applications such as organic light-emitting devices [2] (OLEDs) is no longer a lab curiosity but rather a mature technology for high-performance displays [2,3]. However, solid-state lightning applications that require high brightness are still to be realised due to the insufficient thermal stability of the organic materials. Thermal stress and degradation [4], together with the fact that the luminance and lifetime of OLED devices decreases when operated at high temperatures [5], are widely reported facts. In this respect, an appropriate understanding of thermal transport may help designing materials with tailored heat dissipation characteristics to minimize heat accumulation in OLEDs [6],[7] or to reduce heat flow while increasing charge transport in search for potential thermoelectric applications [8–11]. Their semiconductor nature and the low thermal conductivities make them suitable candidates to improve the thermoelectric figure-of-merit ZT. We note most of the previous studies in this direction have been reported for polymer-based devices and fewer on small molecule organic semiconductors [12].

Physical vapour deposition has been shown to be a suitable tool to tailor the properties of the deposited layers, not only for organic semiconductors [13,14], but also for many other small organic molecules [15–20]. When the deposition conditions, basically substrate temperature and growth rate, are properly set, glasses with exceptional thermodynamic and kinetic stability [13,15,16,21], high densities [14,22–24], low heat capacities [17,25], low water uptake [26] or high moduli [27] can be obtained. These glasses, dubbed ultrastable, are currently gaining widespread attention within the glass community, and a recent report demonstrates the improved packing of these glasses can yield to outstanding improvements in OLED efficiency [28]. An interesting feature of some vapour-deposited organic glasses is that molecules can have average spatial orientations that differ from the random distribution of an isotropic glass. Recent studies have started to focus on the molecular orientation in those materials and its influence on the efficiency of OLEDs [29]. The existence of molecular packing anisotropy in vapour-deposited organic semiconductor thin film glasses was first identified by Lin et al. [30]. Yokoyama and coworkers [31,32] studied the degree of orientation depending on the molecular aspect ratio of the molecule and the deposition conditions. Dalal *et al.* [13] performed dichroism and birefringence measurements on several organic semiconductors, and proposed the ratio between the deposition temperature and the glass transition temperature, $T_{dep}/T_g$, to be the primary parameter affecting the molecular orientation. In particular, it has been shown that the lower the substrate temperature during growth, the higher the tendency towards horizontal orientation. This tunable molecular orientation provides new opportunities to tailor the electrical, thermal and optical properties of the glassy materials.

Many previous studies have focused on the electronic transport properties of organic glasses and crystals, since this is a key parameter for the use of these materials in optoelectronic devices [33]. On the contrary, thermal conductivity measurements in small molecule organic glasses remains largely unexplored and only few studies are reported [34,35]. In general, it is well known that increasing disorder has a remarkable effect on the thermal conductivity. For an inorganic material, such as Silicon, the thermal conductivity varies from 150 W·m$^{-1}$·K$^{-1}$ in bulk Si to around 1.4 W·m$^{-1}$·K$^{-1}$ for the disordered material [36,37]. This low value is frequently understood through the theory of the minimum thermal conductivity where atomic vibrations with mean free paths of the order of the interatomic distance contribute to heat transport [38]. In organic materials the Van der Waals interactions between molecules have a remarkable effect on heat propagation and disorder plays a comparatively less dramatic effect on the thermal conductivity compared to their crystalline counterparts. However, the current understanding of heat conduction in organic glasses is limited by the largely incomplete knowledge about the actual mechanisms ruling over thermal energy exchange in these systems and how the glass atomic-scale morphology affects transport. Measurements on thin-film organic crystals although more abundant also lack a proper understanding of how crystal anisotropy may affect thermal transport along and perpendicular to the molecular chain. The growth of large crystals to minimize the strong influence of grain size on phonon transport is a requirement to unveil the role of crystal anisotropies in heat flow propagation. Ac-calorimetry was previously used to extract the thermal diffusivity of rubrene layers [39]. The thermal anisotropy ratio defined as the relative difference between in-plane and through-plane conductivity, $(k_\parallel - k_\perp)/k_\perp$ was larger than 100% indicating poor thermal transport across the phenyl groups of the rubrene molecules. On the contrary measurements on 6,13-Bis(triisopropylsilylethynyl)pentacene, TIPS-pn, show the through-plane thermal diffusivity is larger than the in-plane one due to an excellent π-orbital overlap [40]. The role of thermal anisotropy has been already addressed in polymeric samples [41] where rubbing or stretching has been used to produce the alignment of the backbone of the polymer along the fibre direction. In this case, the conductivity along the axis of the polymeric chain can be up to 20 times higher than in the perpendicular direction [42].

One difficulty towards a comprehensive analysis of thermal conductivity anisotropies arises from the lack of appropriate experimental techniques to measure the in-plane thermal transport in thin films. Previous in-plane measurements on polymers or polymer nanofibers were conducted on suspended structures [42,43] that directly provide the in-plane thermal conductance. This methodology, frequently used for inorganic nanowires of low thermal conductance, requires lengthy or sophisticated approaches to precisely place the sample bridging the heater/sensor platforms. We have recently shown that a modification of the 3ω–Volklein technique [44–47] can be used to monitor in real-time the growth of organic layers and to measure their in-plane thermal conductance [46]. The high sensitivity of the technique and its versatility makes it an ideal tool to explore the in-plane thermal transport characteristics of organic thin films.

Here we show that by tuning the molecular orientation in glassy films of an organic semiconductor, via variations of the deposition temperature, the thermal anisotropy ratio can be modified to nearly 40%. The achievement of substantial thermal anisotropy in small molecule thin-film glasses is counter-intuitive since structural disorder should lower the anisotropy ratio. By comparing our experimental data with molecular dynamics simulations mimicking the orientation of the vapour-deposited organic layers we disentangle the role of density and molecular orientation on heat propagation. We provide evidence that the change of thermal conductivity is mainly driven by the molecular packing anisotropy in the glass and that thermal transport along the N-N backbone of the N,N'-Bis(3-methylphenyl)-N,N'-bis(phenyl)-benzidine, TPD, molecule is strongly preferred with respect to the perpendicular direction due to a stronger molecular interaction in the former.

**II. METHOS**

**A. Experimental**

*Sample preparation*

A hole transport molecule, N,N'-Bis(3-methylphenyl)-N,N'-bis(phenyl)-benzidine (TPD) was purchased from Cymit Quimica S.L. ( > 99%, purified by sublimation) and used without further purifications. The glass transition (Tg) of the TPD was found to be 333 K, measured by differential scanning calorimetry (DSC) for a sample cooled and heated at 10 K/min. Thin layers of this materials, with thicknesses ranging from 150 nm to 340 nm, were grown by thermal evaporation in a UHV chamber at a base pressure of $5 \cdot 10^{-8}$ mbar using an effusion cell (CREATEC). The deposition rate was monitored using a quartz crystal microbalance (QCM) located close to the substrate and kept constant at 0.21 ± 0.02 nm/s by controlling the effusion cell temperature. The QCM true rate was previously calibrated by measuring a reference sample at the profilometer. The samples were deposited directly onto the silicon nitride membrane-based sensor and subsequently measured in differential mode using the 3ω-Völklein [44–47] method. The sensors, a sample and a reference, were placed on a substrate with a heating element and a Pt100 and the temperature was controlled during the deposition and the measurement of the films. The deposition temperatures ranged from 220 K to 330 K, that is 0.6 to 0.99 times their glass transition temperature. After each deposition, the substrate temperature was always set back to 296 K for the thermal conductivity measurements.

*Thermal conductivity measurements*

*Out-of-plane: 3ω.* The through-plane thermal conductivity was measured with the differential *3ω* method. We used 100 nm thick Al wires with a 5 nm Cr adhesion layer for the *3ω heater/*sensor. The thin film heater of width 5 μm and length 2 mm was defined through the use of a shadow mask during deposition. All measurements were carried out at 290 K in a vacuum of $2.5 \times 10^{-7}$ mbar. Joule heating was produced by applying an AC current to the Al strip using a Keithley 6221 source on the current pads of the sensor. The *3ω* voltage was measured using a Stanford Research 830 lock-in amplifier in the range of 300–3000 Hz. The one-dimensional steady-state heat conduction model was used to extract the cross-plane thermal conductivity values from the temperature rise of the films. Measurements were conducted on samples grown at 220 K and 304 K.

*In-plane measurements: 3ω-Völklein.* We use a modification of the 3ω-Völklein method [46] previously developed by Volklein et al. [47]. In this setup, we use a silicon nitride membrane with the same symmetry as a metal wire that is used for heating/sensing. The sensor is operated in differential mode which allows to measure *in-situ* the conductance with a very high sensitivity $\Delta k/k \simeq 10^{-3}$. We have measured the in-plane thermal conductance of thin films of TPD and α-NPD (shown in SI) deposited over a wide temperature range, from 220 K to 333 K. To obtain the conductance at each $T_{dep}$ we use two different approaches. The first one consisted on single measurements in which 340 nm of TPD were evaporated onto the sensor and subsequently measured. After depositing each film at a given temperature, the sensor was removed from the chamber, cleaned and prepared for the following evaporation. Figure 1 shows (in blue) the thermal conductivity as a function of the deposition temperature obtained using this approach. In the second approach, the thermal conductance was measured at a reference temperature in between depositions at different temperatures of 150 nm thick layers. The layer conductance was, therefore, obtained as the differential before and after each deposition, as sketched in figure 1.

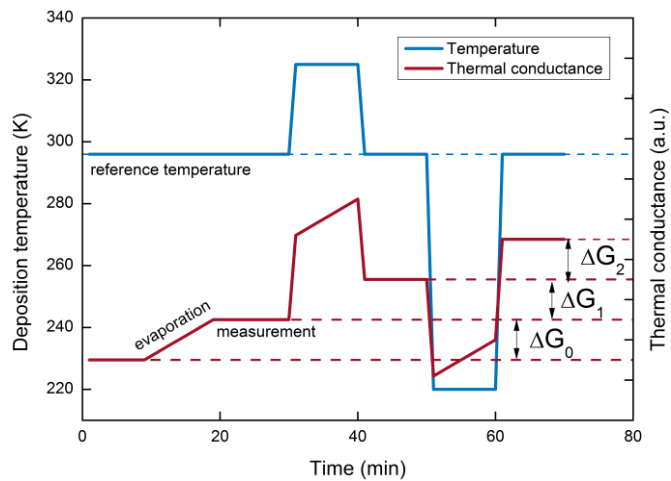

**Figure 1**. **Measurement procedure of the in-plane thermal conductance.** Scheme of the procedure followed to evaluate the thermal conductance of a multilayer stack. The substrate was set to the deposition temperature (in blue) and subsequently cooled or heated at the reference temperature of 295 K. The conductance value is obtained from the differential value before and after the measurement (in red).

**B. COMPUTATIONAL**

*Force fields:* The TPD molecule is modelled according to the CVFF [48] force field, where cross-coupling terms between the various bonded terms provide an accurate description of intra-molecular interactions. The non-bonded interactions are in turn described by a superposition of a Lennard-Jones potential (addressed to describing the van der Waals contribution) and of a Coulomb term as follows

$$\sum \left(\frac{A}{r^{12}} - \frac{B}{r^6}\right) + \sum \frac{q_i q_j}{r}$$

where the sum is performed over all the pairs of non-bonded atoms. The Lennard-Jones term is truncated by a cut-off set at 10.0 Å, while a particle-particle particle-mesh solver approach is adopted to solve the electrostatic problem in the reciprocal space. Coulomb interactions are calculated by assuming fixed charges, as previously obtained by fitting the electrostatic potential of an all-electron Hartree–Fock calculation performed with a medium-sized basis set 6-31G*. The fitting was performed using the RESP method [49,50]. Finally, the Tersoff [51] force field is adopted to describe the silicon substrate, giving such bond-order potential also the role of describing the interactions among molecular carbon atoms and silicon atoms in the substrate.

*Molecular dynamics simulations*: Molecular dynamics simulations are executed by LAMMPS [52] package. Equations of motion are integrated according to the velocity-Verlet integration scheme with a time step of 0.5 fs. Time integration is performed on Nosé-Hoover non-Hamiltonian equations for both constant-pressure (NPT) and constant-temperature (NVT) runs, respectively used to reach equilibrium density at a given temperature and to further equilibrate the sample at a given temperature. Thermostatting and barostatting are achieved by a coupling parameter of 50.0 fs and a relaxation time of 500.0 fs, respectively.

*Sample preparation*: For the anisotropic samples, in order to enforce a preferential molecular orientation parallel to the Si substrate, a layer-by-layer deposition was performed: a first layer of 47 molecules was placed on top of a $20a_0 \times 20a_0$ substrate (where $a_0$=5.4305 Å is the equilibrium silicon lattice constant for Tersoff potential), followed by a geometry optimization and a low temperature (T=1K) annealing for 100 ps. As for *xy-ISO* samples there was no in-plane order, while in the *ANIS* samples molecular axes were mainly oriented along an in-plane direction. This procedure was then repeated piling up to 16 layers: at each step the whole structure was very carefully relaxed. This effectively generated a 6.9 nm-thick film of 752 TPD molecules, which resulted aligned parallel to the substrate. Eventually, the final sample was gently heated up (at $10^{-4}$ K/fs rate) and then equilibrated (400 ps + 100 ps) at the measurement temperature, obtaining a density in the range 1.080-1.085 g/cm$^3$, for the *ANIS* samples, and 1.069-1.079 g/cm$^3$ for the *xy-ISO*. Representative views of the structures are shown in figure S4.

As for the isotropic case, the xy-ISO sample was used as starting configuration: the deposited film was heated up to 900 K during a 300 ps-long NVT run, and then annealed at that temperature for further 500 ps. This allowed the TPD film to completely lose its previous anisotropic structure. We have carefully monitored the molecular orientation on-the-fly to control that a fully isotropic sample was indeed forming. The TPD film was then cooled (at $10^{-4}$ K/fs rate) and the equilibrated at the measurement temperature, resulting into a density in the range 1.059-1.065 g/cm$^3$.

*Calculating thermal conductivity*: The thermal conductivity tensor $\overleftrightarrow{\kappa}$ of a system of volume $V$ is calculated according to the Green-Kubo formalism

$$\kappa_{\alpha\alpha} = \frac{1}{k_B T^2 V} \int_0^\infty \langle J_\alpha(0) J_\alpha(t) \rangle \, dt$$

which relates the ensemble average of the auto-correlation of the heat current vector $\vec{J}$ to the thermal conductivity. The heat current is evaluated through its virial expression [6]

$$\vec{J}(t) = \frac{d}{dt} \sum_i \vec{r}_i(t) \, e_i$$

where $e_i$ is the per-atom energy (potential and kinetic) of the $i$-th particle and $\vec{r}_i$ is the corresponding position. For the here adopted force fields, the evaluation of $\vec{J}(t)$ requires the calculation of a kinetic energy contribution, a pairwise energy contribution, and a reciprocal-space contribution from long-range Coulomb interactions (as well as similar terms for the bond, angle, dihedral and improper energy contributions looped over all the atomic pairs, triplets and quadruplets). In practice: after a 20 ps NVT dynamics, the system is aged in the microcanonical ensemble for another 200 ps, correlating the heat current each 5 timesteps. The values of thermal conductivity are obtained by averaging the resulting $\kappa_{\alpha\alpha}$ over the last 50 ps of dynamics.

## III. RESULTS & DISCUSSION

**Dependence of the thermal conductivity on deposition temperature**

The in-plane thermal conductivity, $k_\parallel$, of the vapour-deposited films was evaluated by the modified 3ω–Volklein technique [44–47], while through-plane, $k_\perp$, measurements were carried out using the 3ω method developed by Cahill and coworkers [53]. The experimentally measured in-plane thermal conductivity of the TPD layers is presented as a function of the deposition temperature in Figure 2a. Similar results obtained with N,N′-bis(1-naphthyl)-N,N′-diphenyl-1,1′-biphenyl-4,4′-diamine (α-NPD) glassy films are shown in supplementary material (SM, Figure S1). Schematics of the 3ω–Volklein and the membrane-based chip used in the measurements are shown in Fig. 2b-d. All measurements were carried out in high vacuum at a fixed temperature of 296 K. The different symbols of Fig. 2a correspond to measurements on various samples following different, but comparable, measurement protocols, as explained in the Methods section. The low values of the in-plane thermal conductivity ≈ 0.16 W·m$^{-1}$·K$^{-1}$ are a preliminary indication of the amorphous character of the layers and that heat is mainly carried by atomic vibrations that, in such systems, are mostly localized (i.e. have quite a short mean free path). $k_\parallel$ decreases first from 0.175 W·m$^{-1}$·K$^{-1}$ (T$_{dep}$ = 220 K, 0.66 T$_g$) to 0.147 W·m$^{-1}$·K$^{-1}$ (T$_{dep}$ = 315 K, 0.94 T$_g$) and then increase to 0.152 W·m$^{-1}$·K$^{-1}$ (T$_{dep}$ = 325 K, 0.98 T$_g$). We note that we were unable to resolve with enough accuracy the thermal conductivity for samples deposited very close to T$_g$ and those points are not shown in Figure 1. The whole dataset is plot in Fig. S2 of SI. The main reason for the lack of reproducibility close to T$_g$ may be related to the high surface mobility and the specific geometry of our sensor that consist on a narrow channel (SiN$_x$ membrane) with sharp Si walls at the edges (see schematics of Figure 2b), conditions that favour the dewetting of the layer from the early stages. In this case, the uncertainty in the amount of mass sensed produces low accuracy and reproducibility of the thermal conductivity. At temperatures below 325 K the measurements at different temperatures were reproducible with an uncertainty of ±0.004 W·m$^{-1}$·K$^{-1}$. The major sources of uncertainty are coming from the determination of the evaporated thickness and the geometrical dimensions of the heater/sensor.

Interestingly enough, the dependence of the in-plane thermal conductivity with the deposition temperature is in line with several previous studies providing evidence that the properties of vapour-deposited organic glasses can be tailored by tuning the deposition conditions [13,16,22,54]. In particular, the highest densities and thermal and kinetic stabilities are achieved at deposition temperatures in the vicinity of 0.85 T$_g$. TPD has maxima in stability and density at 285 K (0.85 T$_g$), above and below this temperature both the density and the stability decrease [14]. Typical calorimetric traces for TPD glasses are plotted in Fig. S3 showing the enhanced stability. Films grown at around 0.85 T$_g$ show both an increase of the onset of devitrification, T$_{on}$, during heating scans and a larger overshoot at T$_{on}$, clear signatures of their enhanced kinetic and thermodynamic stability.

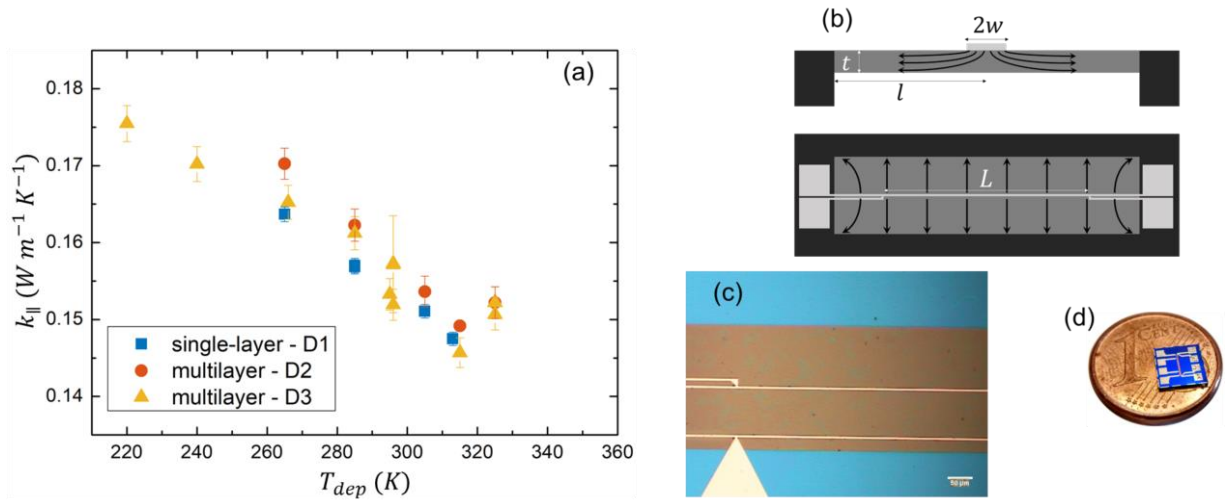

**Figure 2. Heater/sensors for 3ω-Volklein thermal conductivity and measured values for TPD**. (a) Thermal conductivity versus deposition temperatures in TPD glasses. Data points correspond to two different measurement procedures as further explained in Methods. Blue-square points were determined from single independent evaporations and cleaning the sensor after each measurement. Triangles and circles correspond to data obtained with a continuous method using two different devices, in which multilayers are deposited sequentially and the differential conductance provides values for each individual layer. (b) Schematic design of the sensor and the direction of the heat flow. (c) optical images of membrane (brown) and the heater/sensor (central metallic line) and upper view of the chip (d).

In order to understand if the observed variation in $k_\parallel$ correlates to changes in the density of the glassy films we plot in Fig. 3 this variable obtained from ref. [14] (red-circles, left axis) together with the in-plane thermal conductivity (blue-squares, right axis) versus $T_{dep}$. It is readily apparent from this figure that density and thermal conductivity do not correlate over the substrate temperatures explored here. While the denser glasses are obtained at $T_{dep}$ ≈ 0.80-0.85 Tg, the maximum value of $k_\parallel$ is reached at 0.66 Tg (ca. 220 K). In fact, an increase in density should be followed by an increase in thermal conductivity and not the contrary, as clearly occurs in the region 0.65-0.85 Tg. The inset in Fig. 3 sheds light on this assumption by plotting the relative thermal conductivity, determined from Green Kubo molecular dynamic simulations, versus the density of several isotropic glasses spanning more than 15% in density variation. The formation of these simulated glasses is explained in the Methods and shown in Figure S4, though we remark that the molecular unit is built with the consistent valence force field (CVFF) to take account of intra-molecular interactions and by a superposition of a Lennard-Jones potential and a Coulomb term for the non-bonded interactions. From this data, a 1.5 % increase in density should produce an increase of the thermal conductivity by around 2.5 %. On the contrary, the experimental data show that a glass grown at 220 K with a density lower by 0.4 % with respect to a glass grown at 315 K undergoes an increase of the thermal conductivity by 15 %. The lack of correlation between thermal conductivity and density indicates that there might be some other factor accounting for the thermal conductivity variation with $T_{dep}$.

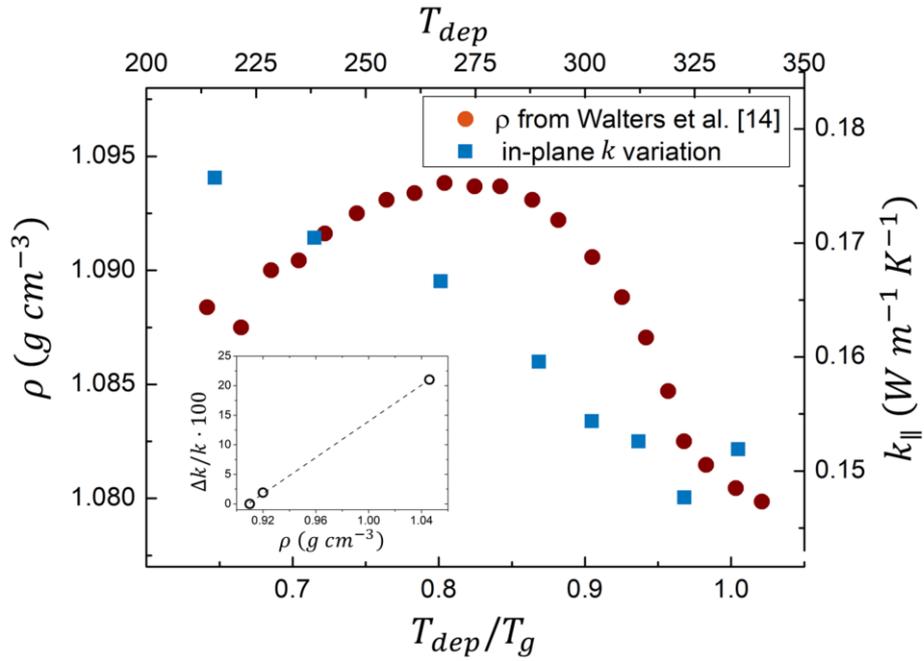

**Figure 3. Thermal conductivity compared to density of the stable glasses.** Thermal conductivity and density variation in TPD as a function of deposition temperature relative to the glass transition temperature. Density data taken from Walters et al. [14] assuming a density of 1.08 g/cm² for the conventional liquid. inset: MD simulations of the role of density in the thermal conductivity of an isotropic glass.

**Influence of molecular packing**

Figure 4 shows the relative variation of $k_\parallel$ (black squares, left-axis) normalized to the 325 K value, 0.152 ± 0.004 W·m$^{-1}$·K$^{-1}$. In red circles (right axis), the order parameter $S_z$ extracted from Dalal *et al.* [13] is plotted in the same figure with a common x-axis. The order parameter $S_z$ is a measure of the average orientation of the molecules. It is defined as $S_z = (3 <cos^2\theta_z> -1)/2$, where $\theta_z$ is the angle of the long molecular axis relative to the substrate normal. Its definition is such that the $S_z$= -0.5, 0, or 1.0 value is, respectively, indicating that all the molecules are oriented parallel to the plane of the substrate, a totally random orientation, or a perfect vertical alignment. The correlation between this parameter and the thermal conductivity variation is remarkably good as seen from Fig. 4. When the molecules are in preference horizontally oriented (T$_{dep}$ < 290 K), the in-plane thermal conductivity is higher. It is worth noticing that even the small peak of $S_z$ at 315 K (0.95 Tg), indicating a small tendency to vertical orientation, is reproduced in the thermal conductivity data. Our results suggest that heat transport in the parallel direction is favoured when molecules tend to lie with the long molecular axis (along the N-N axis) parallel to the surface (see sketches in figure 4). On the contrary, in-plane thermal transport is sizeably reduced when molecules tend to align along the out-of-plane direction. At first glance this is somewhat surprising given the small size of the molecular unit and the disorder inherent to the glass that should minimize the influence of molecular orientation. To confirm the existence of thermal anisotropy we carried out through-plane thermal conductivity measurements with the 3ω technique on two TPD samples, one grown at

$T_{dep}$ = 220 K with the molecules preferentially aligned parallel to the substrate and another at $T_{dep}$= 304 K with isotropic orientation. The results are summarized in Fig. 5. Since we were mostly interested on the relative variation of the thermal conductivity between both samples we did not carry thickness dependent measurements to remove the influence of the thermal interface resistance between dissimilar layers. Since both samples have the same thickness and similar thermal interface barriers the relative difference is accurate enough to provide a suitable description. For the sample grown at 220 K we got a $k_\perp$ 22 % lower than in the isotropic film. This result agrees with the previous in-plane measurements (Fig. 3) and the influence of the molecular packing anisotropy since thermal transport in the perpendicular direction is low when molecules lie roughly parallel to the substrate, i.e. when $k_\parallel$ is maximal. Therefore, considering that the isotropic sample has identical parallel and perpendicular conductivities we obtain a value of the thermal anisotropy ratio measured at a T = 220 of ≈ 37 % (Figure 5).

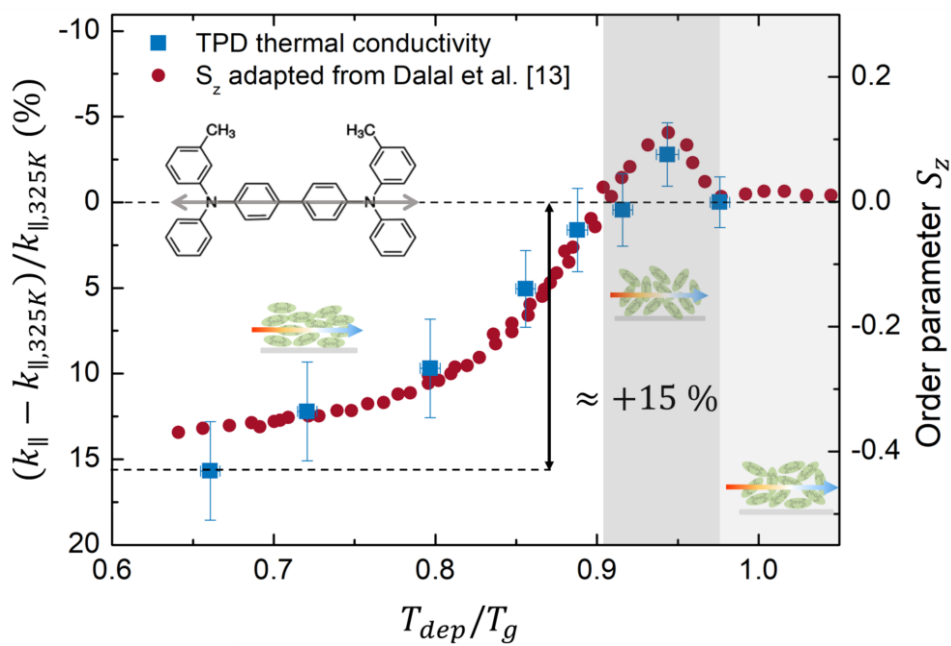

**Figure 4**. **Correlation between thermal conductivity and molecular orientation**. Comparison between the relative variation of the thermal conductivity and the order parameter obtained from Dalal et al. [13] plotted against the deposition temperature scaled by Tg.

In order to underpin the effect of the molecular packing anisotropy on thermal transport in organic glasses we carried out extensive molecular dynamics simulations on a system with tailored orientation of the TPD molecules (see Fig. S4 for more details about the structure).

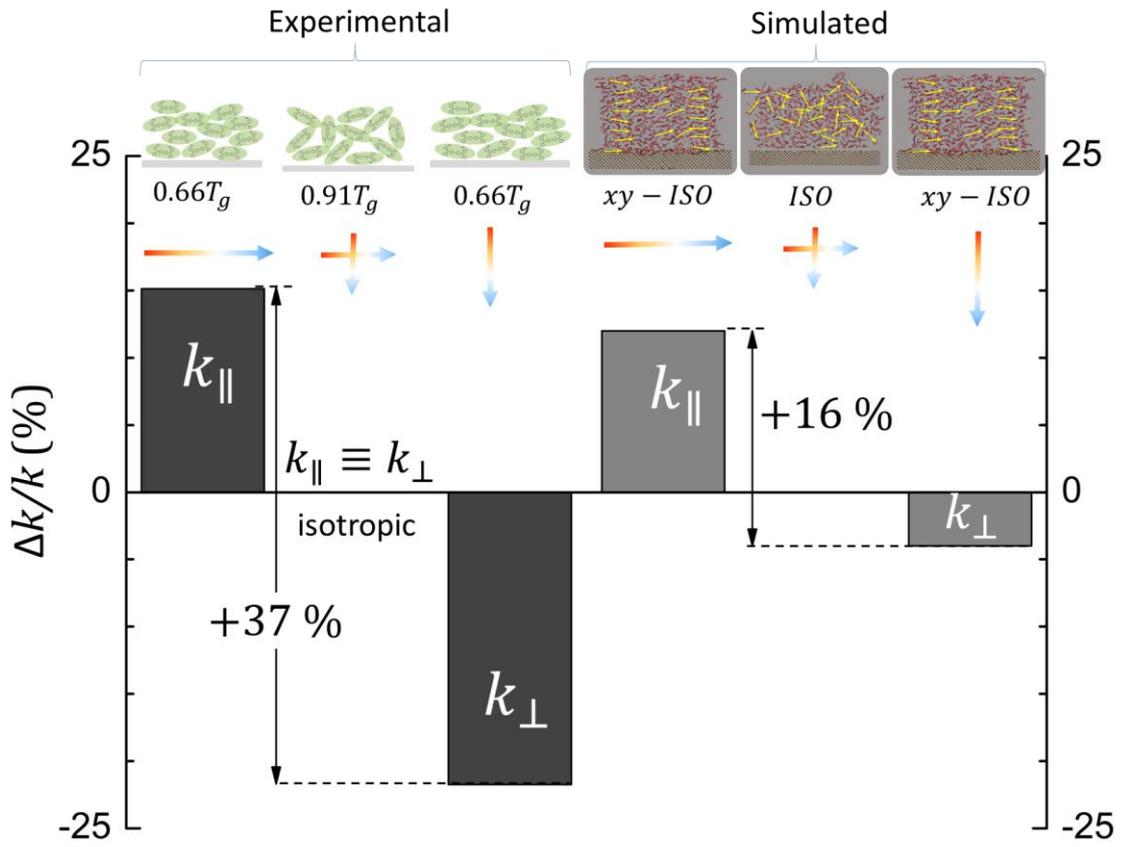

**Figure 5**. **Thermal anisotropy ratio for experimental and simulated samples.** Schematics showing the relative variation of thermal conductivity for the various experimental (left) and simulated (right) samples. Sample grown at 220 K (ca. 0.66 Tg) is roughly equivalent to the simulated xy-iso. The upper panel illustrates the structure of the experimental and simulated samples. The coloured arrows indicate the direction of the heat flux. The yellow arrows in the simulated samples indicate the long molecular axis.

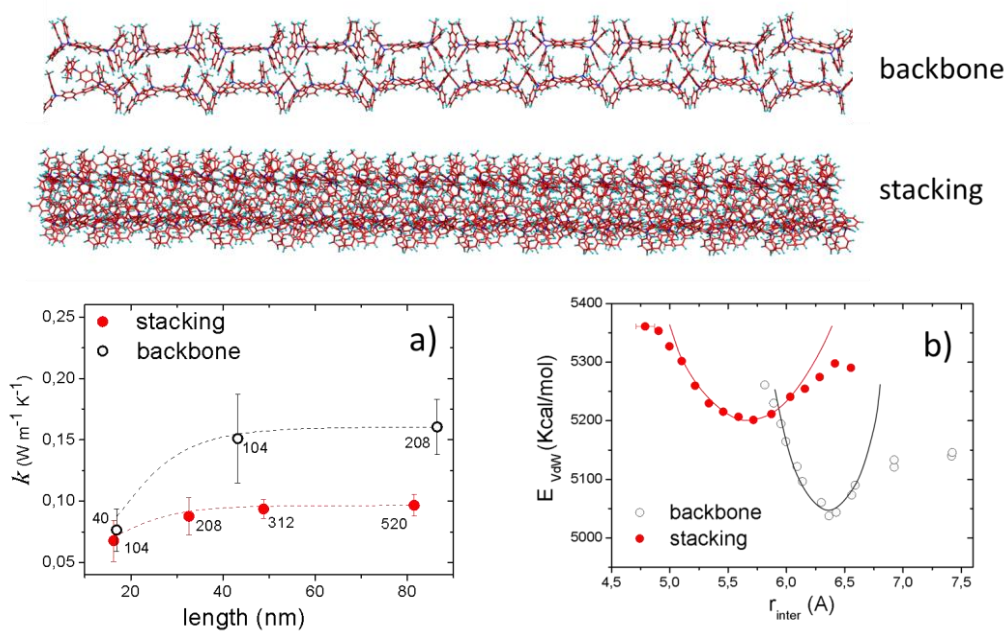

**Figure 6. Thermal conductivity and Van der Waals energy of simulated quasi-1D structures.** Upper panel: backbone and stacking quasi-1D structures. Lower panel: (a) thermal conductivity versus length for the quasi-1D structures. (b) Configurational energy due to Van der Waals interactions versus intermolecular spacing.

A set of computational samples (Figure 5 upper panel and Figure S4) with density values similar to the experimental ones (See SM for details) and different packing arrangements was built to mimic the low-temperature vapour-deposited ones. Specifically, we generated three unlike samples differing for their main molecular orientation. The *ISO* sample was obtained simply by quenching from the melt and, therefore, it results fully isotropic or, equivalently, it does not show any preferential alignment of TPD molecular axes. The remaining two samples, respectively referred to as *xy-ISO* and *ANIS*, were both characterized by a preferential in-plane molecular orientation, where the reference plane is in any case the substrate surface: this feature has been obtained by enforcing the planar alignment of molecular axis during the sample preparation. However, the two systems differ in that the in-plane alignment is totally random in the *xy-ISO* sample or further enforced to align the *x* direction for the *ANIS* sample, respectively, In short, the sequence (*ISO*)-(*xy-ISO*)-(*ANIS*) provides an increasing character of molecular anisotropy. A quantitative analysis of the molecular packing orientation in the simulated samples and the calculated X-ray diffraction pattern is presented in SM. The relative difference between the thermal conductivity of the *xy-ISO* sample in the in-plane direction and the corresponding through-plane value is 16 %, in qualitative agreement with the experimental data, as show in Figure 5. Sample *ANIS* shows an even higher thermal anisotropy ratio between $k_{\parallel,x}$ (in-plane along x) and the through-plane thermal conductivity of 95 %.

Our results (both modelling and experimental data) indicate that the variations of thermal conductivity are mainly driven by the molecular packing anisotropy in the glass and that thermal transport along the N-N backbone of the TPD molecule is preferred with respect to transport along the direction perpendicular to the phenyl rings. We propose that this effect could be interpreted in terms of a different efficiency in transmitting heat carriers, depending on the direction of the heat flow with respect to the molecular orientation. In order to substantiate this conjecture, we simulated two different quasi-1D structures, namely: (i) a linear bundle of TPD molecules aligned along the N-N axis and (ii) a line-up of TPD molecules aligned along the direction of π-π bonding. Hereafter we will refer to such configurations as backbone stacking or π-π stacking, respectively. They are shown in the upper panel of Fig. 6 together with a plot of the thermal conductivity versus length for the linear chains, Fig. 6a. The overall trend shows a saturation for increasing length and confirms that the direction perpendicular to the backbone is detrimental for heat transport. Present simulations intelligibly reported a 70 % higher thermal conductivity for the backbone configuration. In order to explain this result, we have calculated the interaction strength between molecules in both stacking arrangements as function of the inter-molecular spacing. More specifically, the average intermolecular distance <$r_{inter}$> was varied in the range 4.75 Å ≤ $r_{inter}$ ≤ 7.5 Å and the corresponding configurational energy has been computed as shown in Figure 6b. Such a potential energy nearby the equilibrium distance is basically harmonic, while at smaller/higher distances the onset of anharmonic behavior is observed, as expected. A parabolic fitting near the minimum yields the effective force constants characterizing the intermolecular coupling within a simple spring-and-ball picture. The quadratic fit for both set of data is a function of the form $f(x) = K(x - x_0)^2 + C$, where K is the spring constant and C is the energy minimum located at a $x_0$. The obtained values are $K_{backbone}$ =824.7 kcal/mol·Å², $X_{0,backbone}$=6.39 Å and $C_{backbone}$=5047.2 kcal/mol, for the backbone configuration and $K_{stacking}$=312.7 kcal/mol·Å², $X_{0,stacking}$=5.67 Å and $C_{stacking}$=5202.8 kcal/mol for the stacking configuration. Consistently with the adopted picture, we argue a stiffer effective spring value translates into a more efficient thermal conduction, according to the following twofold heuristic argument. In general, thermal conductivity is proportional to the group velocity of heat carriers: since a larger force constant causes a steeper vibrational branch, this reflects into a higher group velocity. On the other hand, we understand that heat current basically represents the energy transferred by a flux of carriers corresponding to atomic vibrations: the higher the force constant, the higher the vibrational energy, the higher the energy of such heat carriers. We note that the thermal anisotropy reported here qualitatively agrees with observations in aligned polymers systems where an increase in the strength of intermolecular forces leads to an enhancement of the thermal conductivity [42].

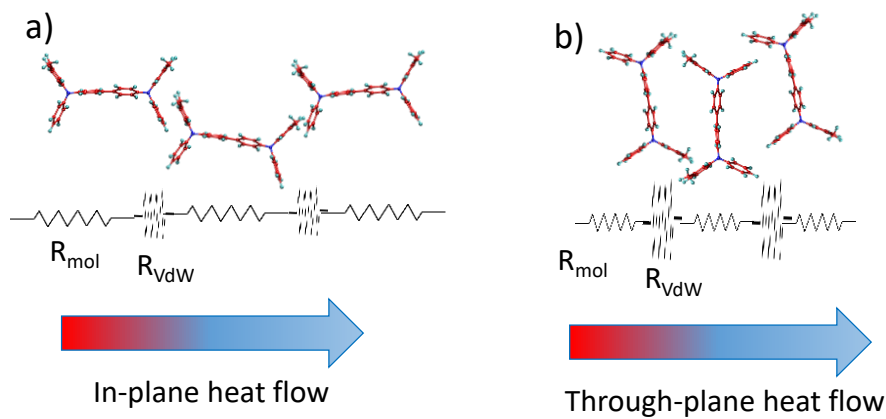

**FIGURE 7**. **Thermal resistive network**. Schematics of thermal resistance network in the in-plane direction (a) and the through-plane direction (b). $R_{VdW}$ accounts for the strength of the intermolecular interactions and $R_{mol}$ represents the intramolecular thermal resistance.

The in-plane and through-plane conductance can be estimated considering that the molecules and VdW interactions form a thermal resistive network, as shown in Figure 6. We assume that the interface thermals resistances (ITR) that correspond somehow to the coupling between neighbouring molecules that are joined through weak Van der Waals interactions dominate thermal transport. A stronger molecular interaction is represented by a lower *ITR* (higher thermal interface conductance, TIC) and the intermolecular π-π stacking entails a higher thermal resistance between molecules. According to Fourier's law the heat flux, Q, is proportional to the temperature difference *ΔT*, *Q=G·ΔT*, where *G* is the thermal conductance. If the series *ITR* dominate heat transfer, G will be equal to the thermal interface conductance due to VdW interactions and the thermal conductivity may be simply written as $k = L \cdot G_{VdW}$, where *L* is the average distance between molecules (see SM for details about the calculation). To estimate values of in-plane and through-plane $G_{VdW}$ we take the average distance between molecules evaluated through the X-ray diffraction patterns of the simulated samples (see SM for more details about the X-ray profiles). The obtained values $L_\parallel = 6.0 \pm 0.2$ Å and $L_\perp = 4.58 \pm 0.2$ Å are in excellent agreement with experimental data measured by Gujral et al. [55] Using the experimental values of the thermal conductivities, $k_\parallel = 0.175\ W\ m^{-1}K^{-1}$ and , $k_\perp = 0.110\ W\ m^{-1}K^{-1}$, gives $G_{VdW,\parallel} \approx 300\ MW\ m^{-2}K^{-1}$ and $G_{VdW,\perp} \approx 230\ MW\ m^{-2}K^{-1}$. We note that given the inherent disorder of the samples the evaluated conductance should be considered as an average contribution to the thermal boundary conductance of the various VdW interactions between different entities of nearest-neighbour molecules and therefore cannot be directly compared to the directional force constants evaluated previously. Although a direct comparison to previous TIC values on other systems is problematic due to the impact of disorder on our data we provide a comparison to previous works. Our values are significantly lower than those reported for metal/dielectric interfaces, approximately $1\ GW\ m^{-2}K^{-1}$ and comparable to the calculated TIC between different crystallographic orientations in crystalline dinaphtho[2,3-b:2',3'-f ]thieno[3,2-b]thiophene (DNTT) that ranges from 150-300 $MW\ m^{-2}K^{-1}$ [56] or to the interface between myoglobin proteins that amounts to 301 $MW\ m^{-2}K^{-1}$ at 320 K [57]. The data for TPD lies between those of organic–organic interfaces such as copper phthalocyanine (CuPc)–fullerene (C60) interfaces, (TIC≈400 $MW\ m^{-2}K^{-1}$) and organic/inorganic interfaces such as pentacene/metal (TIC ≈ $10\ MW\ m^{-2}K^{-1}$) [58] or CuPc-Au (TIC ≈ $20\ MW\ m^{-2}K^{-1}$) that is purely a VdW-like interaction [59].

## 4. CONCLUSION

In summary, we have observed that the in-plane thermal conductivity of vapour-deposited thin film stable glasses of TPD strongly depend on the deposition temperature and a thermal anisotropy ratio of ≈ 40 % is achieved at a T$_{dep}$ of 220 K (0.66 T$_g$). At this temperature molecules are on average located with the phenyl rings perpendicular to the growth direction and there is a strong anisotropy between the in-plane and through-plane directions. This packing anisotropy has a strong effect in heat propagation. The microscopic details of heat transport are revealed by molecular dynamics simulations that show the different strength of the molecular interaction in the direction along the backbone of the molecules compared to the direction of the π-π stacking. The stronger molecular interaction along the backbone favours the propagation of microscopic heat carriers along this direction. This is at odds with electronic transport that favours propagation along the perpendicular direction to the long axis of the molecule provided there is sufficient π-π interaction. This strategy could be employed in future developments to implement small molecule thin films for its use in thermoelectric-based applications.

**ACKNOWLEDGMENTS**

This work was financially supported by the Spanish MINECO MAT2013-40896-P and MAT2016-79579-R together with the European Regional Development Funds (ERDF). P. Ferrando-Villalba and J. Ràfols-Ribé are also in receipt of a FPU grant from the Spanish Ministry of Education, Culture and Sport at the time of the study. Ll. Abad thank the MINECO for a Ramón y Cajal Contract (RYC-2013-12640) and CSIC for JAE-INTR_14_00479.

# Evidence of thermal transport anisotropy in stable glasses of vapour deposited organic molecules

Joan Ràfols-Ribé[1], Riccardo Dettori[2], Pablo Ferrando-Villalba[1], Marta Gonzalez-Silveira[1], Ll. Abad[3], Aitor Lopeandía[1], Luciano Colombo[2], Javier Rodríguez-Viejo[1,*]

[1]Nanomaterials and Microsystems Group, Physics Department, Universitat Autònoma de Barcelona, 08193 Bellaterra, Spain.

[2]Department of Physics, University of Cagliari, Cittadella Universitaria, 09042 Monserrato (Ca), ITALY.

[3]IMB-CNM-CSIC, Campus Bellaterra, 08193 Bellaterra, Spain

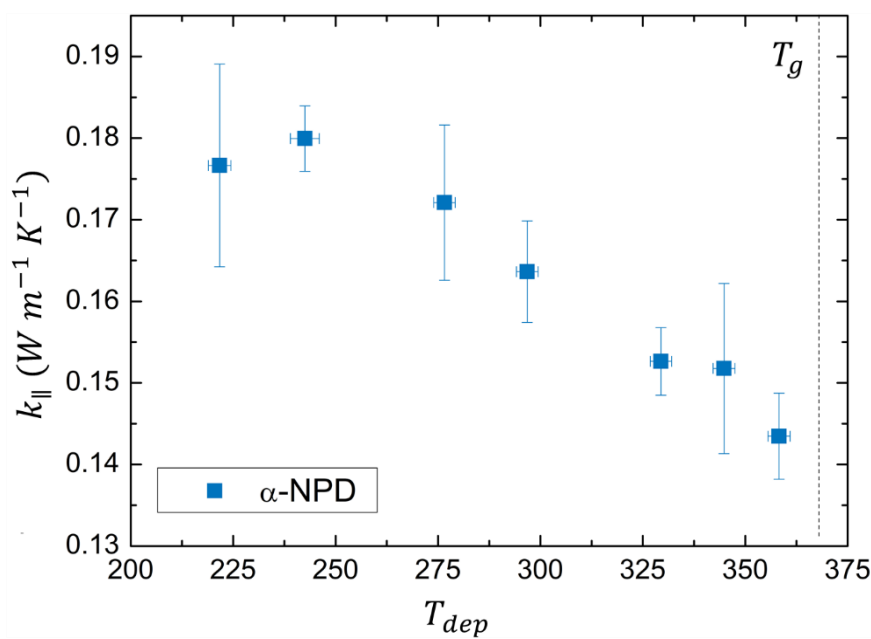

*Figure S1*. Thermal conductivity versus deposition temperatures in α-NPD thin films glasses. Data points correspond to the average $k_\parallel$ value obtained from two to three measurements at each temperature using the multilayer approach and 150 nm thick layers (see Methods). Error bars correspond to the standard deviation (2σ).

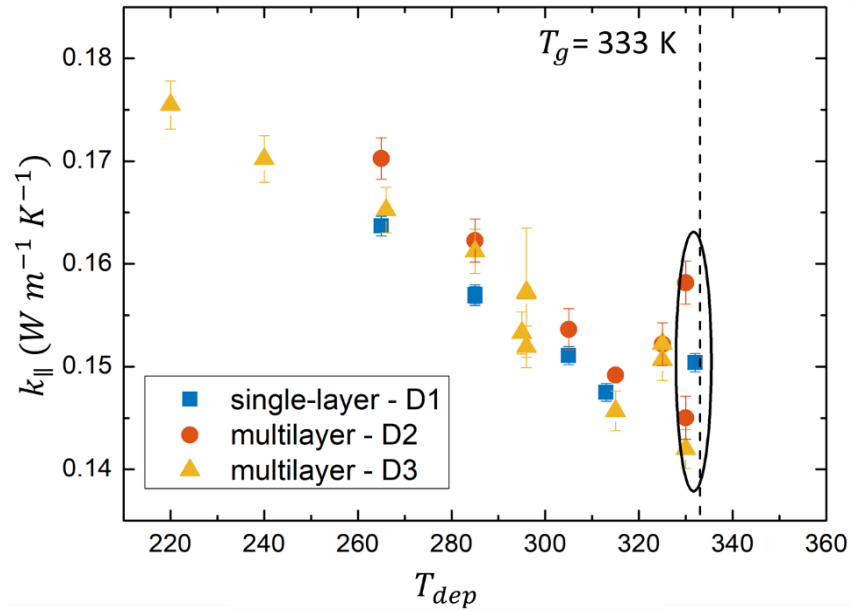

*Figure S2*. Same thermal conductivity data for TPD showed in figure 1 but including also the non-reproducible measurements near the glass transition temperature at 333 K.

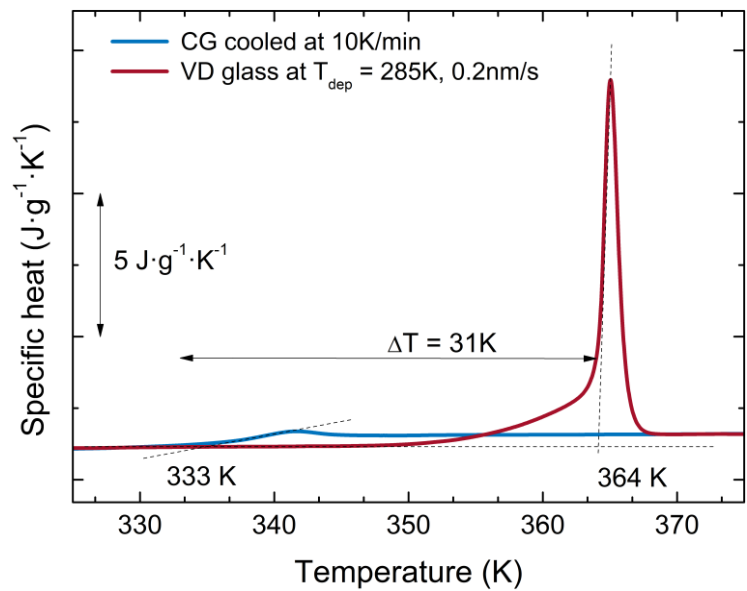

*Figure S3*. Calorimetric trace of the devitrification of three differently prepared glasses of TPD. CG stands for conventional glass, which is the glass prepared from the liquid at a cooling rate of 10 K/min and VD stands for vapor-deposited, indicating also the deposition temperature ($0.86T_g$) and growth rate. The dashed lines indicate the determination of the onset temperatures.

**Simulated structures**

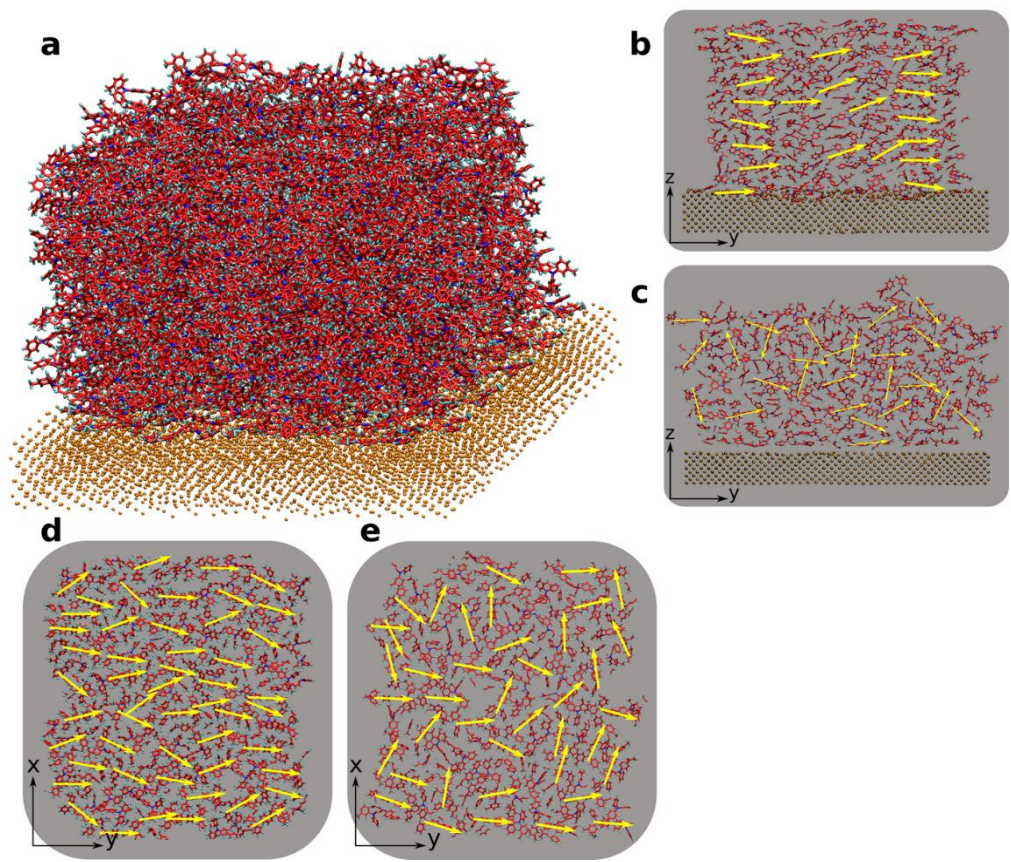

*Figure S4.* Panel a: perspective view of the simulated TPD film deposited on Si substrate; panels b and c**:** side view of, respectively, xy-ISO and ISO samples; panels d and e: top view of, respectively, ANIS and xy-ISO samples. Yellow arrows represent the orientation of the molecule backbone, reflecting the anisotropy of the systems.

**Density of simulated layers**

*2.1 Density*
All samples have been very carefully relaxed so as to get them at the (nearly) experimental density values. Results are reported in the following table

| Sample | Density (g/cm$^3$) |
|--------|---------------------|
| ISO    | 1.059---1.065       |
| xyISO  | 1.069---1.079       |
| ANIS   | 1.080---1.085       |

The reported values are very close to experimental samples. In particular, ANIS and ISO differ of about 1.5---2.0%, while xy---ISO and ISO differ of about 0.5---1.4%.

**Molecular orientation in the simulated samples**

The spatial orientation of TPD molecules can be unambiguously specified by defining two vectors, namely: (i) the vector identifying the N-N direction of the molecule and (ii) the vector identifying the plane of the molecule. This latter is defined by considering the vector product between the two vectors joining one nitrogen atom to two carbon atoms of the opposite aromatic ring of the TPD backbone (Fig. S5).

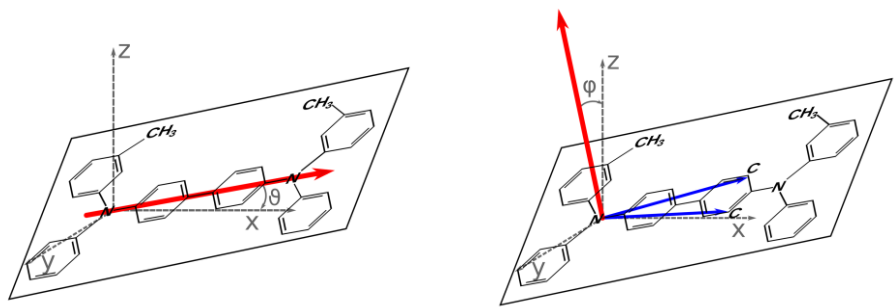

*Figure S5*. (left)Projection of the NN vector along x,y,z directions and (right) orientation of the molecular plane.

By computing the distribution of the angles that vectors (i) and (ii) form with the three directions (x, y, z) will provide informations on the overall molecular orientations in the sample. In Fig. S6 a thorough comparison between the ANIS and xyISO samples is performed by considering the orientation of the NN vector (panels a, b and c); the ANIS and ISO samples are in turn compared by considering the orientation of the molecular plane (panels d, e and f). All data are calculated at T=280K. We can extract quite a few information. The ANIS sample is highly peaked for $\cos(\theta)=\pm 1$ which explains the higher efficiency of thermal transport along x direction. On the other hand, the xyISO sample shows a much flatter distribution due to its random orientation in the xy-plane (see panel a). Furthermore, the ANIS sample shows a (broad) distribution around $\cos(\theta)=0$, which is a fingerprint of a preferential normal orientation of the molecules with respect to the y direction. Again, for the xyISO sample, the distribution is much flatter and contains non-zero values for a greater range of $\cos(\theta)$ values (see panel b). Both samples show a distribution peaked around $\cos(\theta)=0$, confirming the π-π stacking along the z direction: this reflects in a less efficient thermal transport (see panel c). Another interesting features is that ANIS sample shows a distribution peaked around $\cos(\Phi)=0$, suggesting that molecules are on average parallel to the xy-plane and aligned along x. The ISO sample shows a broad and flatter distribution due to a random orientation of the molecular plane (see panel d). The ANIS sample also show a flat distribution, that is a random orientation of the plane with respect to y (not perfectly parallel to y). This does not affect thermal transport since for such a system thermal conduction preferentially occurs along the x direction. The ISO sample has the same distribution (see panel e). Finally, the distribution for z component (describing the orientation with respect the substrate plane) is peaked at $\cos(\Phi)=\pm 1$ for the ANIS sample, accounting for a preferential stacking arrangement along the z direction. The same distribution for the ISO sample, in turn, is almost flat, as a consequence of the isotropy of the system. The small shoulders at $\cos(\Phi)=\pm 1$ are relative to the molecules directly attached to the substrate and to the molecules in the last plane of the TPD film, since they preserve an overall planar arrangement (see panel f).

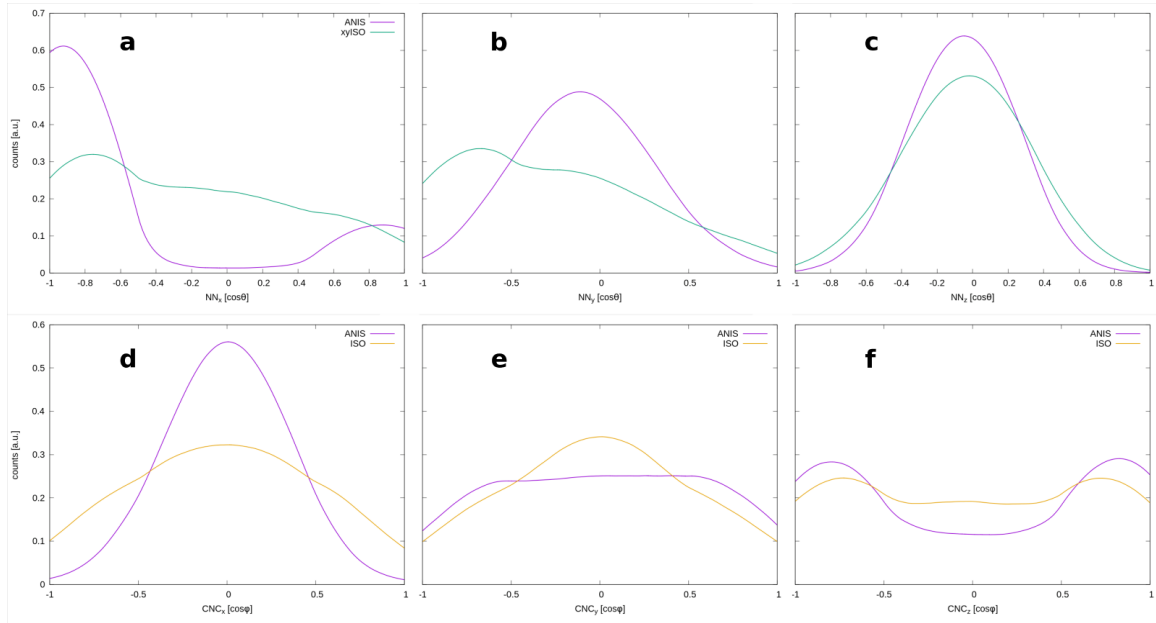

**Figure S6.** Upper panel: Orientation of the NN vector for the ANIS and xyISO samples. Lower panel: Orientation of the molecular plane for the ANIS and ISO samples.

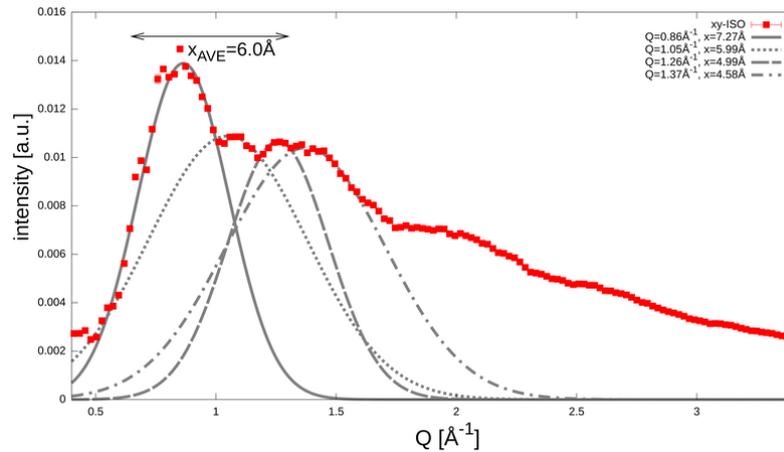

**Figure S7.** XRD pattern of the xy-iso simulated sample. 6 Å is the average value of intermolecular distances in the xy-plane.

**In-plane and through-plane conductance**

We assume the molecule and de VdW interactions can be seen as a thermal resistive network. Therefore the total thermal resistance can be written as the sum of the two series resistances (see Figure SX) $R_T = R_{T,mol} + R_{T,VdW} = N_{mol} \cdot R_{mol} + N_{VdW} \cdot R_{VdW}$, where $N_{mol}$ and $N_{VdW}$ are the number of molecular and VdW units and $R_{mol}$ and $R_{VdW}$ the individual resistances associated to each unit, respectively. Writing the above eq. in terms of the individual conductivities

$$R_T = N_{mol} \cdot \frac{L_{mol}}{k_{mol} \cdot A} + N_{VdW} \cdot \frac{L_{VdW}}{k_{VdW} \cdot A}$$

With $L_{mol}$ and $L_{VdW}$ being the lengths of the units.

Therefore, the thermal conductance

$$G_T = k_{meas} \cdot \frac{A}{L} = (R_T)^{-1} = \frac{k_{mol} \cdot k_{VdW} \cdot A}{k_{mol} \cdot N_{VdW} \cdot L_{VdW} + k_{VdW} \cdot N_{mol} \cdot L_{mol}}$$

$L$ is the total thickness and $k_{meas}$ is the measured thermal conductivity.

To simplify if we consider that the conductance is dominated by an interface thermal resistance due to VdW interactions, we can write $k_{mol} \gg k_{VdW}$ that leads to a simplified expression

$$\frac{k_{meas}}{L} = \frac{k_{VdW}}{N_{VdW} \cdot L_{VdW}} = \frac{G_{VdW}}{N_{VdW}} \Rightarrow G_{VdW} = k_{meas} \cdot \frac{N_{VdW}}{L} = k_{meas}/(L_{mol} + L_{VdW})$$

$G_{VdW}$ stands for the conductance of an individual unit (the interface thermal conductance due to VdW interactions). The average distance between molecules ($L_{mol} + L_{VdW}$) in the in-plane and through-plane directions is calculated from the simulated X-ray diffraction profiles (Figure S7) and compared to the values derived by Gujral et al. [S1] that used X-ray diffraction to evaluate the structure of a TPD sample deposited at 260 K. i.e. with molecules on average oriented parallel to the substrate surface. The simulated average distances are 4.58 Å and 6.0 Å for the through-plane (z plane) and in-plane (xy plane) respectively. Experimentally the distance between molecules in the through-plane direction is well defined by the low-angle XRD peak located at q ≈1.4 A$^{-1}$ that gives 4.5 A. The value associated to the xy-plane is more difficult to evaluate since several low-angle peaks or shoulders appear at 1.4 A$^{-1}$, 1.2 A$^{-1}$ and 0.75 A$^{-1}$. A rough averaging gives a mean distance value around 6 Å. The agreement between simulated and experimental data support the suitability of the simulated structure.